\documentclass[prb,twocolumn,amsmath,amsymb]{revtex4}

\usepackage{graphicx}
\usepackage{dcolumn}
\usepackage{bm}

\begin{document}

\title{The influence of structural short-range order on the phase diagrams \\ of diluted FCC magnet with arbitrary spin \\and modified RKKY interaction
}

\author{Karol Sza{\l}owski}
\email{kszalowski@uni.lodz.pl}
\author{Tadeusz Balcerzak}%
\affiliation{%
Department of Solid State Physics, University of \L\'{o}d\'{z},\\
ulica Pomorska 149/153, 90-236 \L\'{o}d\'{z}, Poland
}%

\date{\today}

\begin{abstract}
A diluted FCC magnet with modified long-range RKKY interaction and arbitrary Ising spin $S$ is considered within two-sublattice model. In the molecular field approximation the Gibbs free-energy is derived, from which all magnetic thermodynamic properties can be self-consistently obtained. In particular, the phase diagrams are studied for different magnetic ion and free-charge concentration, the atomic short-range-order (Warren-Cowley) parameter being taken into account.
\end{abstract}

\maketitle

\section{Introduction}

The diluted magnetic systems constitute an important field of research in solid-state and statistical physics. The description of such alloys involves different challenges, starting from the question of the electronic structure of inhomogenous systems lacking translational symmetry through the calculation of exchange interactions between spins and finally to the problem of magnetic ordering. Another issue is the usual difference in the energy scales of magnetic and non-magnetic interactions between the ions in the alloy. Studies of such substances have recently attracted attention owing to the interest in diluted magnetic semiconductors (DMS), important from the spintronics point of view.~\cite{ohnoscience}    

The vital importance of disorder in description of DMS has recently been emphasised by numerous theoretical works and stimulated by the experimental data showing the great sensitivity of the critical temperature and magnetic ordering in Ga$_{x}$Mn$_{1-x}$As samples to their treatment even though the concentration of magnetic impurities remains unchanged.~\cite{potashnik} This suggests that the occupation of the lattice sites by magnetic ions may not be completely random.   

For the case of this celebrated DMS, Ga$_{x}$Mn$_{1-x}$As, the effects of magnetic impurity clustering on the critical temperature of a model DMS have been investigated; for instance,  by means of MC method in the work of Priour and Das Sarma~\cite{dassarmaclustering}, who found only a weak influence of magnetic ion aggregating on the phase transition. By contrast, Bouzerar \emph{et al.}~\cite{bouzerar} predicted a noticeable increase in Curie temperature of the clustered impurities system for Ga$_{x}$Mn$_{1-x}$As and Ga$_{x}$Mn$_{1-x}$N. The presence of clustering in DMS has been supported theoretically, for example, by the first-principle calculations of various alloy characteristics performed by Drchal \emph{et al.}~\cite{drchal}, Kudrnovsk\'{y} \emph{et al.}~\cite{kudrnovskyphase} or Raebiger \emph{et al}.~\cite{raebiger} On the other hand, numerous existing works use different approaches to the clustering in order to show that it decreases the critical temperature in Ga-based DMS.~\cite{xu,tang,franceschetti,berciu}

Various variants of \emph{ab initio} methods, yielding the crystalline, electronic and magnetic structure, are believed to provide reliable estimations of critical temperatures for specific substances (for example see Ref.~\onlinecite{tc}). At the same time, however, it is difficult to extract from those methods the systematic analysis of the effect of particular factors on the final parameters of the system. Here we see the importance  of simplified, schematic models which could be extensively analysed, providing some detailed insight, for instance, into the general features of the influence of the disorder on the magnetic properties.
Our aim is to analyze a model diluted magnet with modified RKKY interaction, focusing on the importance of structural correlations in distribution of magnetic moments on the lattice. We improve the virtual crystal approximation by taking into consideration the correlations of pairs and study its effect on the magnetic ground state phase diagrams as well as the critical temperature. Moreover, we introduce a measure of magnetic frustration and discuss its sensivity to structural correlations.  

The paper is organized as follows: In IInd section the theoretical model is described in detail by a statistical-thermodynamical method. In particular, the analytical expressions for the Gibbs energy and phase transition temperature are derived. On this basis the numerical calculations are carried out in IIIrd section and the results are presented in figures. A recently developed approach~\cite{szalowskibalcerzak} of numerical summation over arbitrary large number of co-ordination zones is adopted. Discussion of the results in IIIrd sec. is focused on the influence of atomic dilution and short-range-order on the magnetic phase diagrams. Finally, in IVth section some conclusions are drawn. The paper is supplemented by Appendix containing a method of configurational averaging in the pair approximation.


\section{Theoretical model}

We will consider the Ising-type Hamiltonian on the diluted FCC lattice containing antiferromagnetic nearest neighbour (NN) interaction as well as the long-range indirect interaction of the RKKY kind. In order to describe various antiferromagnetic structures a model of two interpenetrating sublattices $(a,b)$ is adopted. The Hamiltonian can be written in the form:
\begin{eqnarray}
\mathcal{H}&=&\!-\sum_{\left\langle i,j\right\rangle}^{}{\!J_{ij}\,\xi_i\xi_j\,S^a_iS^a_j}
\!-\sum_{\left\langle i,j\right\rangle}^{}{\!J_{ij}\,\xi_i\xi_j\,S^b_iS^b_j}
\nonumber\\
&&\!-\sum_{\left\langle i,j\right\rangle}^{}{\!J_{ij}\,\xi_i\xi_j\,S^a_iS^b_j}
-h\sum_{i}^{}\xi_{i}S^{a}_{i}-h\sum_{i}^{}\xi_{i}S^{b}_{i}\nonumber\\
\label{eq1}
\end{eqnarray}
where $S^{\alpha}_{i}=-S,...\, ,+S$ is the Ising spin of arbitrary magnitude $S$ situated in $i$-th lattice site and belonging to the sublattice $\alpha$ $(\alpha=a,b)$. In Eq.~(\ref{eq1})
$h=-g^{\mathrm {eff}}\mu_{\mathrm B}H^{z}$ corresponds to the external magnetic field $H^{z}$ oriented in $z$-direction, whereas $g^{\mathrm {eff}}$ is the effective gyromagnetic factor, which for the case of RKKY interaction has been introduced in Ref.~\onlinecite{balcerzak1}. The occupation operators $\xi_{i}=(0,1)$ describe the magnetic dilution. Namely, $\xi_{i}=0$ corresponds to the magnetic vacancy in the $i$-th lattice site, $\xi_{i}=1$
 describes the state when $i$-th lattice site is occupied by the spin $S^{\alpha}_{i}$. These operators are subject to configurational averaging $\left<...\right>_{r}$, which we assume to be independent of the magnetic structure. On the other hand, the spin operators $S^{\alpha}_{i}$, for a given atomic configuration, are subject to thermal averaging $\left<...\right>$ only.\\

The configurational averaging of single-site occupation operators can be conveniently described by introducing a parameter $n=\left<\xi_{i}\right>_{r}$, where $n$ is a concentration of magnetic component.  The parameter $n$ can be regarded as a quotient of the number of magnetic atoms to the total number of lattice sites. For simplicity, we will further assume that $n$ is equal for both sublattices $(\alpha =a,b)$. In turn, the configurational averaging of $(\xi_{i}\xi_{j})$-pairs leads to the expression:
\begin{equation}
\left<\xi_{i}\xi_{j}\right>_{r}^2 =n^2+ \Delta_{k}
\label{eq2}
\end{equation}
where
$\xi_{i}=\left<\xi_{i}\right>_{r} +\delta \xi_{i}$, and $\Delta_{k}=\left<\delta \xi_{i}\delta \xi_{j}\right>_{r}$ is a fluctuation of the occupation numbers, which is characteristic for the $k$-th coordination zone. As has been shown in the Appendix, the fluctuations must obey the sum rule:
\begin{equation}
\sum_{k}^{} z_{k} \Delta_{k}=0
\label{eq3}
\end{equation}
where $z_{k}$ is the total number of lattice sites on the $k$-th co-ordination zone where the fluctuation $\Delta_{k}$ takes place. These fluctuations are connected with the Warren-Cowley (W-C) short-range-order (SRO) parameter $\alpha_k$ by the relationship:
\begin{equation}
\alpha_{k}= \frac{\left<\xi_{i}\xi_{j}\right>_{r}-\left<\xi_{i}\right>_{r}\left<\xi_{j}\right>_{r}}{\left<\xi_{i}\right>_{r}\left<\xi_{j}\right>_{r}} =\frac{\Delta_{k}}{n^{2}}
\label{eq4}
\end{equation}
A detailed analysis of the physical range of W-C parameter resulting from the pair probability distribution $p\,(\xi_{i}\xi_{j})$ is presented in the Appendix.\\

As far as the thermal averaging of the spin operators is concerned, we will adopt the simplest molecular field approximation (MFA), with the decoupling relation $\left<S^{\alpha}_{i} S^{\beta}_{j}\right>\approx m^{\alpha} m^{\beta}$ where $\alpha$ (or $\beta$)$=a,b$ and $m^{\alpha}=\left<S^{\alpha}_{i}\right>$ denotes $\alpha$-sublattice magnetization. The MFA is justified both by the presence of long-range interaction (for the infinite interaction range it becomes an exact method) and by the sublattice model of antiferromagnetism according to the idea of N\'{e}el. As far as we know,~\cite{anderson,smart,morrish} in the FCC structure, apart from the ferromagnetic (F) and paramagnetic (P) phases, nothing impedes the notion that different antiferromagnetic orderings exist. The most known seem to be the antiferromagnetic 1st kind (AF1), antiferromagnetic 1st kind improved (AF1I) and 2nd kind antiferromagnetic (AF2) orderings.\\

Within MFA the magnetic enthalpy can be found by the configurational and thermal averaging of the Hamiltonian (\ref{eq1}). The result is:
\begin{eqnarray}
H=\left<\left<\mathcal{H}\right>\right>_{r}&=&\!-\frac{N}{4}\sum_{k}^{}{\!J_{k}z^{\uparrow\uparrow}_{k}\left(n^2 + \Delta_{k}\right)\,\left[\left(m^a\right)^{2}+\left(m^b\right)^{2}\right]}\nonumber\\
&&\!-\frac{N}{2}\sum_{k}^{}{\!J_{k}z^{\uparrow\downarrow}_{k}\left(n^2 + \Delta_{k}\right)\,m^a m^b}\nonumber\\
&&\!-\frac{N}{2}\, n\, \left(m^a +m^b\right)h
\label{eq5}
\end{eqnarray}
By $N$ we denote the total number of lattice sites, whereas the summation upon $k$ is performed over all co-ordination zones centered at the arbitrary lattice site. In Eq.~(\ref{eq5}) $z^{\uparrow\uparrow}_{k}$ ( $z^{\uparrow\downarrow}_{k}$ ) are the number of lattice sites on the $k$-th co-ordination zone, whose spins (if occupied) are oriented parallelly (antiparallelly) to the central spin. Thus, $z^{\uparrow\uparrow}_{k}$ is the co-ordination number at the $k$-th zone formed from lattice sites belonging to the same magnetic sublattice as the central spin, while $z^{\uparrow\downarrow}_{k}$ is the co-ordination number formed from lattice sites belonging to different sublattice. Those numbers satisfy the condition $z^{\uparrow\uparrow}_{k}+z^{\uparrow\downarrow}_{k}=z_{k}$ and their disribution upon $k$ depends on the type of magnetic ordering (F, AF1, AF1I or AF2).\\

The exchange integral $J_{k}$ in Eq.~(\ref{eq5}) for a given co-ordination zone $k$ is basically the RKKY long-range interaction, with the exception of the first co-ordination zone, where we additionally include the antiferromagnetic superexchange interaction $J^{\,\mathrm{AF}}<0$. This kind of interaction has been introduced in several papers~\cite{dietlscience,dassarma} concerning diluted magnetic semiconductors (DMS). Thus, we assume that $J_{1}= J^{\,\mathrm{AF}}+J^{\,\mathrm{RKKY}}_{1}$ for $k=1$ and $J_{k}=J^{\,\mathrm{RKKY}}_{k}$ for $k=2,3,...$, where the RKKY interaction is given by the expression:~\cite{rudermankittel}
\begin{equation}
J^{\,\mathrm{RKKY}}_{k}=C\left(k_{\mathrm{F}}a\right)^4\frac {\sin\left(2k_{\mathrm{F}}r_{k}\right)
-2k_{\mathrm{F}}r_{k}\cos\left(2k_{\mathrm{F}}r_{k}\right)}{\left(2k_{\mathrm{F}}r_{k}\right)^{4}}\,
e^{-r_{k}/\lambda}
\label{eq6}
\end{equation}
In Eq.~(\ref{eq6}) $a$ is the lattice constant and $r_{k}$ stands for the radius of the $k$-th co-ordination zone. 
The Fermi wavevector $k_{\mathrm{F}}$ for FCC structure takes the form $k_{\mathrm{F}}=\left(12\pi^2n\right)^{1/3}/a$, where we assume that each occupied lattice site yields one charge carrier to the conduction or valence band. Thus, we assume that the free carriers concentration is $n$, i.e., the same as the concentration of magnetic atoms in the FCC lattice. The energy constant $C$ in Eq.~(\ref{eq6}) can be treated as the unit energy, both for the exchange integral and for the $k_{\mathrm{B}}T$ scale as well. As far as the exponential factor in Eq.~(\ref{eq6}), containing the damping parameter $\lambda$, is concerned, such a term has been introduced by Mattis~\cite{mattis} in order to account for the charge carriers localization in disordered systems. It follows from the literature,~\cite{degennes,kudrnovskyrkky,dassarmaclustering,dassarma} that such localization takes place in some DMS, such as Ga$_{1-x}$Mn$_x$As. The above description of $J_{k}$ for $k=1,2,...$ defines the so-called modified RKKY interaction, the model which has successfully been used in several papers concerning DMS systems.~\cite{dassarma}\\

The magnetic enthalpy (\ref{eq5}) allows the studies of the ground-state phase diagrams (for $T \to 0$) when a perfect spin alignment is assumed (for instance, $m^{a}=S$ and $m^{b}=-S$ for the antiferromagnetic phase). By comparison of the enthalpy values for different magnetic phases (F, P, AF1, AF1I and AF2) the stability areas for each phase can be established from the minimum condition. The magnetic phase diagram can be obtained vs. concentration $n$, fluctuation distribution $\Delta_{k}$ and external field $h$, for given parameters $J^{\,\mathrm{AF}}$ and $\lambda$ characterizing the modified RKKY interaction. For instance, for the antiferromagnetic phases in the ground state the enthalpy per lattice site is given by the expression:
\begin{equation}
\frac{H}{N}=-\,\frac{1}{2}\,n^2S^2\sum_{k}^{}\left(1+\alpha_{k}\right){\left(z^{\uparrow\uparrow}_{k}-z^{\uparrow\downarrow}_{k}\right)}\,J_{k}
\label{eq7}
\end{equation}
where $\alpha_{k}$ is W-C parameter for the $k$-th co-ordination zone and $z^{\uparrow\uparrow}_{k}$  ($z^{\uparrow\downarrow}_{k}$) depend on the type of the antiferromagnetic phase. On the other hand, for the ferromagnetic phase in the ground state (when $m^{a}=m^{b}=S$)  the corresponding formula for the enthalpy reads:
\begin{equation}
\frac{H}{N}=-\,\frac{1}{2}\,n^2S^2\sum_{k}^{}\left(1+\alpha_{k}\right)z_{k}\,J_{k} + \frac{1}{2}\, nSh
\label{eq8}
\end{equation}
In turn, for the paramagnetic phase (with $m^{\alpha}=0$) we assume $H/N=0$. The numerical calculations of the ground state phase diagrams with the structural ordering taken into account will be presented in the next section.\\

As far as the temperature studies are concerned, one has to consider not only the temperature dependencies of the sublattice magnetizations $m^{\alpha}$ ($\alpha=a,b$) but also the magnetic entropy. In the MFA method a unified approach can be suggested, based on the single-site density matrix
\begin{equation}
\rho^{\alpha}_{i}=\,\frac{\mathrm{exp}\left[\beta\left(\Lambda^{\alpha}+h\right)S^{\alpha}_{i}\right]}{Z^{\alpha}_{1}}
\label{eq9}
\end{equation}
where $\beta=1/k_{\mathrm{B}}T$, and $\Lambda^{\alpha}$ is a variational parameter of the molecular field acting on $\alpha$-sublattice. The single-site partition function $Z^{\alpha}_{1}$ for $\alpha$-sublattice is defined by the formula:
\begin{equation}
Z^{\alpha}_{1}=\,\mathrm{Tr}_{i}\{\mathrm{exp}\left[\beta\left(\Lambda^{\alpha}+h\right)S^{\alpha}_{i}\right]\}
=\sum_{l=-S}^{S}\mathrm{exp}\left[\beta\left(\Lambda^{\alpha}+h\right)l\right]
\label{eq10}
\end{equation}
Then, the total partition function $Z$ in the MFA is given by the product: $Z=\left(Z^{a}_{1}Z^{b}_{1}\right)^{Nn/2}$, where $Nn/2$ is the number of lattice sites occupied within one sublattice. With the use of single-site density matrix (\ref{eq9}) the various thermal mean values can be calculated. For instance, the magnetization of the occupied site on $\alpha$-sublattice is:
\begin{equation}
m^{\alpha}=\mathrm{Tr}_{i}\left[S^{\alpha}_{i}\rho^{\alpha}_{i}\right]=\frac{1}{Z^{\alpha}_{1}}\mathrm{Tr}_{i}\{S^{\alpha}_{i}\mathrm{exp}\left[\beta\left(\Lambda^{\alpha}+h\right)S^{\alpha}_{i}\right]\}
\label{eq11}
\end{equation}
which leads to the general formula
\begin{equation}
m^{\alpha}=SB\left(S\beta\left(\Lambda^{\alpha}+h\right)\right)
\label{eq12}
\end{equation}
where $\alpha=a,b$ and $SB\left(x\right)$ is the Brillouin function for an arbitrary spin $S$:
\begin{equation}
SB\left(Sx\right)=\frac{2S+1}{2}\,\mathrm{coth}\left(\frac{2S+1}{2}\,x\right) - \frac{1}{2}\,\mathrm{coth}\left(\frac{x}{2}\right)
\label{eq13}
\end{equation}\\

Due to the factorization in MFA the total entropy $\sigma$ can be presented as a sum of single-site entropies for both sublattices:
\begin{equation}
\sigma=\frac{N}{2}\,n\sum_{\alpha=a,\,b}^{}\sigma^{\alpha}_{1}
\label{eq14}
\end{equation}
The single-site entropy $\sigma^{\alpha}_{1}$ for the occupied site on $\alpha$-sublattice is given by the thermal mean value
\begin{equation}
\sigma^{\alpha}_{1}=\,-k_{\mathrm{B}}\left<\mathrm{ln}\rho^{\alpha}_{i}\right>=\,-k_{\mathrm{B}}
\mathrm{Tr}_{i}\left(\rho^{\alpha}_{i}\mathrm{ln}\rho^{\alpha}_{i}\right)
\label{eq15}
\end{equation}
Making use of Eq.~(\ref{eq9}), the single-site entropy $\sigma^{\alpha}_{1}$ can be presented in the form:
\begin{equation}
\sigma^{\alpha}_{1}=\,-\frac{1}{T}\left(\Lambda^{\alpha}+h\right)m^{\alpha}+k_{\mathrm{B}}\mathrm{ln}Z^{\alpha}_{1}
\label{eq16}
\end{equation}
hence the total entropy (\ref{eq14}) can be presented as:
\begin{eqnarray}
\sigma&=&\frac{N}{2}\,n\frac{1}{T}\left[ -\left(\Lambda^{a}+h\right)m^{a}-\left(\Lambda^{b}+h\right)m^{b}\right. \nonumber\\ &&+ \left. k_{\mathrm{B}}T\mathrm{ln}Z^{a}_{1}+k_{\mathrm{B}}T\mathrm{ln}Z^{b}_{1}\right]
\label{eq17}
\end{eqnarray}\\

Having calculated the entropy (\ref{eq17}) and the enthalpy (\ref{eq5}), the Gibbs free-energy can be found in MFA from the thermodynamic formula:
\begin{equation}
G=H-\sigma T
\label{eq18}
\end{equation}
However, with a view to constructing the phase diagrams, we are interested in the Gibbs energy per lattice site, i.e., in the chemical potential $\mu$, which is given by:
\begin{eqnarray}
\mu=\,\frac{G}{N}&=&-\frac{1}{4}n^2\sum_{k}^{}{\!J_{k}z^{\uparrow\uparrow}_{k}\left(1 + \alpha_{k}\right)\,\left[\left(m^a\right)^{2}+\left(m^b\right)^{2}\right]}\nonumber\\
&&\!-\frac{1}{2}n^2\sum_{k}^{}{\!J_{k}z^{\uparrow\downarrow}_{k}\left(1 + \alpha_{k}\right)\,m^a m^b}\nonumber\\
&&\!+\frac{n}{2}\left(\Lambda^{a}m^{a}+\Lambda^{b}m^{b}\right) - \frac{n}{2}k_{\mathrm{B}}T\left(\mathrm{ln}Z^{a}_{1}+\mathrm{ln}Z^{b}_{1}\right)\nonumber\\
\label{eq19}
\end{eqnarray}
The molecular field (variational) parameters $\Lambda^{\alpha}$ appearing in Eqs.~(\ref{eq19}), (\ref{eq12}) and (\ref{eq10}) can be determined from the necessary extremum conditions:
\begin{equation}
\frac{\partial \mu}{\partial \Lambda^{\alpha}}=0
\label{eq20}
\end{equation}
(for $\alpha=a,b$) which lead to the following expressions:
\begin{equation}
\Lambda^{a}=\sum_{k}^{}{\!J_{k}\left(1 + \alpha_{k}\right)\,\left( z^{\uparrow\uparrow}_{k} m^a +z^{\uparrow\downarrow}_{k}m^b\right)}
\label{eq21}
\end{equation}
and
\begin{equation}
\Lambda^{b}=\sum_{k}^{}{\!J_{k}\left(1 + \alpha_{k}\right)\,\left( z^{\uparrow\uparrow}_{k} m^b +z^{\uparrow\downarrow}_{k}m^a\right)}
\label{eq22}
\end{equation}
Now, with the help of Eqs.~(\ref{eq21}) and (\ref{eq22}) the chemical potential (\ref{eq19}) for the thermodynamical equilibrium is obtained in the final form:
\begin{eqnarray}
\mu&=&\frac{1}{4}n^2\sum_{k}^{}{\!J_{k}z^{\uparrow\uparrow}_{k}\left(1 + \alpha_{k}\right)\,\left[\left(m^a\right)^{2}+\left(m^b\right)^{2}\right]}\nonumber\\
&&\!+\frac{1}{2}n^2\sum_{k}^{}{\!J_{k}z^{\uparrow\downarrow}_{k}\left(1 + \alpha_{k}\right)\,m^a m^b}\nonumber\\
&&\!-\frac{n}{2}k_{\mathrm{B}}T\left(\mathrm{ln}Z^{a}_{1}+\mathrm{ln}Z^{b}_{1}\right)
\label{eq23}
\end{eqnarray}
together with $Z^{\alpha}_{1}$ given by Eq.~(\ref{eq10}) and $m^{\alpha}$ as a solution of Eq.~(\ref{eq12}).\\

From this point on, the expression (\ref{eq23}) for the chemical potential allows the self-consistent studies of all the thermodynamic properties in MFA. 
Let us notice first that the necessary equilibrium conditions for the chemical potential (\ref{eq20}) are also satisfied with respect to the sublattice magnetization: ${\partial \mu}/{\partial m^{\alpha}}=0$ ($\alpha=a,b$). Moreover, the mean magnetization per one lattice site $m$ can be derived alternatively to Eqs.~(\ref{eq11}, \ref{eq12}) merely by differentiation of the chemical potential over the external field:
\begin{equation}
m=\left(\frac{\partial \mu}{\partial h}\right)_{T}=\,\frac{1}{2}n\left(m^{a}+m^{b}\right)
\label{eq24}
\end{equation}
Analogously, by differentiation of the chemical potential over temperature the mean entropy per lattice site can be calculated, yielding the same form as Eqs.~(\ref{eq14},\ref{eq17}):
\begin{equation}
\frac{\sigma}{N}=\left(\frac{\partial \mu}{\partial T}\right)_{h}=\,\frac{1}{2}n\left(\sigma^{a}_{1}+\sigma^{b}_{1}\right)
\label{eq25}
\end{equation}
 Consequently, other thermodynamic properties such as the magnetic susceptibility, or magnetic contribution to the specific heat can be calculated as the second-order derivatives of the chemical potential (\ref{eq23}).\\
 \begin{figure*}
\includegraphics[scale=0.85]{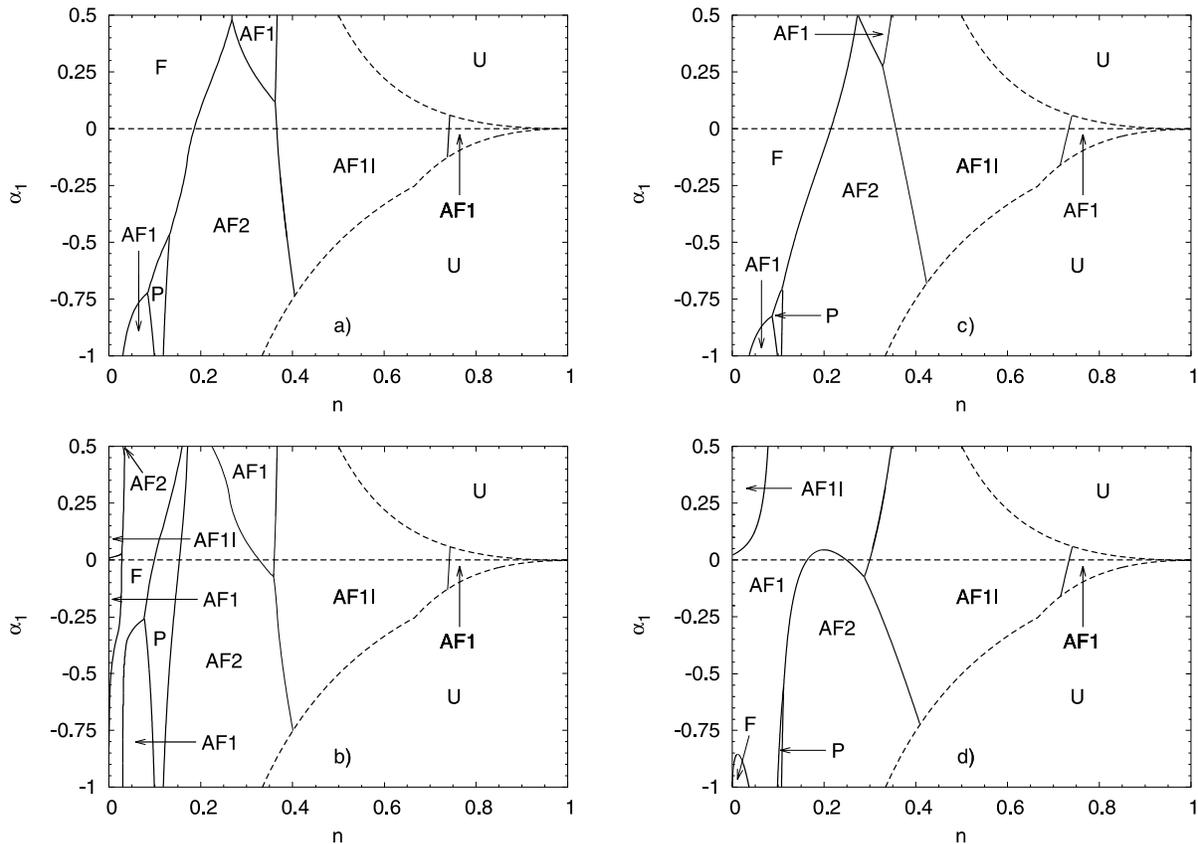}
\caption{\label{fig:fig1}Ground-state magnetic phase diagrams in the ($n, \alpha_1$)-plane for different values of parameters modifying the RKKY interaction: $\lambda\to\infty$, $J^{\mathrm{AF}}=0$ (a), $\lambda\to\infty$, $J^{\mathrm{AF}}/C=-0.5$ (b), $\lambda/a=1.0$, $J^{\mathrm{AF}}=0$ (c), $\lambda/a=1.0$, $J^{\mathrm{AF}}/C=-0.5$ (d). By U we denote an unphysical area for $\alpha_1$-parameter.}
\end{figure*}
 The critical temperature of the second-order (continuous) phase transitions can be obtained from the linearization of Eq.~(\ref{eq12}) for $h=0$ and $m^{\alpha}\to0$. 
 Making use of the linear expansion for the Brillouin function $SB\left(Sx\right)\stackrel{\left(x \to 0\right)}{\longrightarrow} S\left(S+1\right)x/3$ we obtain from Eq.~(\ref{eq12}):
\begin{equation}
m^{\alpha}=\frac{S\left(S+1\right)}{3}\beta_{\mathrm{c}}\Lambda^{\alpha}
\label{eq26}
\end{equation}
($\alpha=a,b$), where $\beta_{\mathrm{c}}=1/k_{\mathrm{B}}T_{\mathrm{c}}$ and $T_{\mathrm{c}}$ is the critical temperature. Now substituting $\Lambda^{\alpha}$ from Eqs.~(\ref{eq21}) and (\ref{eq22}) into (\ref{eq26}), we obtain a set of two linear, homogeneous equations for $m^{\alpha}\to0$ in the vicinity of $T_{\mathrm{c}}$. Next, by setting the determinant to be equal zero, the phase transition temperature $T_{\mathrm{c}}$ is derived in the following form:
\begin{equation}
k_{\mathrm{B}}T_{\mathrm{c}} = \frac{S\left(S+1\right)}{3}\,n\sum_{k}^{}{\!J_{k}\left(1 + \alpha_{k}\right)\,\left( z^{\uparrow\uparrow}_{k} \pm z^{\uparrow\downarrow}_{k}\right)}
\label{eq27}
\end{equation}
Eq.~(\ref{eq27}) is a generalization of MFA result for the long-range RKKY interaction with the structural clustering taken into account.
The solution with "$+$" corresponds to the Curie temperature and is applicable to ferromagnetic phase transition, whereas the solution with "$-$" corresponds to the N\'{e}el temperature for the antiferromagnetic (AF1, AF1I, and AF2) phase boundaries. The temperature phase diagrams based on Eq.~(\ref{eq27}) will be calculated in the next section.\\

\section{Numerical results and discussion}

The numerical studies have been carried out for the model FCC structure with dilution and the modified RKKY interaction taken into account. The external magnetic field $h=0$ was set. According to the theoretical considerations, when the dilution is not random it cannot be desribed by only a single variable $n$, but should be characterized by the set of Warren-Cowley SRO parameters. These parameters, $\alpha_k$, which are incorporated into the analytical formulas in previous section, fulfil the sum rule (Eq.~\ref{eq3}) and are treated as independent of the magnetic structure. However, for the simplicity of numerical calculations, we will further assume that only $\alpha_1$ and $\alpha_2$ parameters are different from zero, while $\alpha_k=0$ for $k>2$. Such assumption reflects the empirical fact that the structural correlations for two first coordination zones are the  most important factors. In this way only $\alpha_1$ becomes an independent SRO parameter of the theory (with some constraints imposed, as discussed in the Appendix (see \ref{eq:ineq2} and Fig.~6)), whereas $\alpha_2$ is determined from the sum rule: $\alpha_2=-\left(z_1/z_2\right)\alpha_1$.\\

All numerical calculations involving summation over co-ordination zones $k$ have been performed up to $k_{\mathrm{max}}=41253$, which corresponds to the radius of interaction $r_{k_{\mathrm{max}}}/a=150$ in the FCC structure. The co-ordination numbers $z^{\uparrow\uparrow}_{k}$  ($z^{\uparrow\downarrow}_{k}$) for a particular magnetic phase have been computed by a computer program analysing the lattice. For such large range of interaction a perfect numerical convergency of the summation has been achieved. Hence, for the results presented below, the numerical errors are negligible. In particular, in case of the ground-state phase diagrams, where no analytical approximations have been made, the presented results can be regarded as near-exact.\\

\subsection{Ground-state phase diagrams}
The ground-state phase diagrams have been computed on the basis of Eqs.~(\ref{eq7}) and (\ref{eq8}) for $h=0$. 
The stability regions for the phases (F, P, AF1, AF1I and AF2), where the enthalpy of each phase is in minimum, have been found in ($n, \alpha_1$)-plane, and the results are presented in  Fig.~\ref{fig:fig1}. The (a)-(d) parts of  Fig.~\ref{fig:fig1} correspond to the different parameters of modified RKKY interaction, as indicated in the figure caption. In all parts (a)-(d) the unphysical (U) area of Warren-Cowley parameter $\alpha_1$ vs. charge carriers concentration $n$ is depicted and delimitated by the dashed lines. The origin of that area was discussed in the Appendix. The horizontal dashed line for $\alpha_1=0$ corresponds to te absence of SRO. The ocurrence of phases and their sequence for  $\alpha_1=0$ upon $n$ is in agreement with those studied in Refs.~\onlinecite{szalowskibalcerzak,balcerzak2}, where the SRO was not taken into account.\\
\begin{figure*}
\includegraphics[scale=0.82]{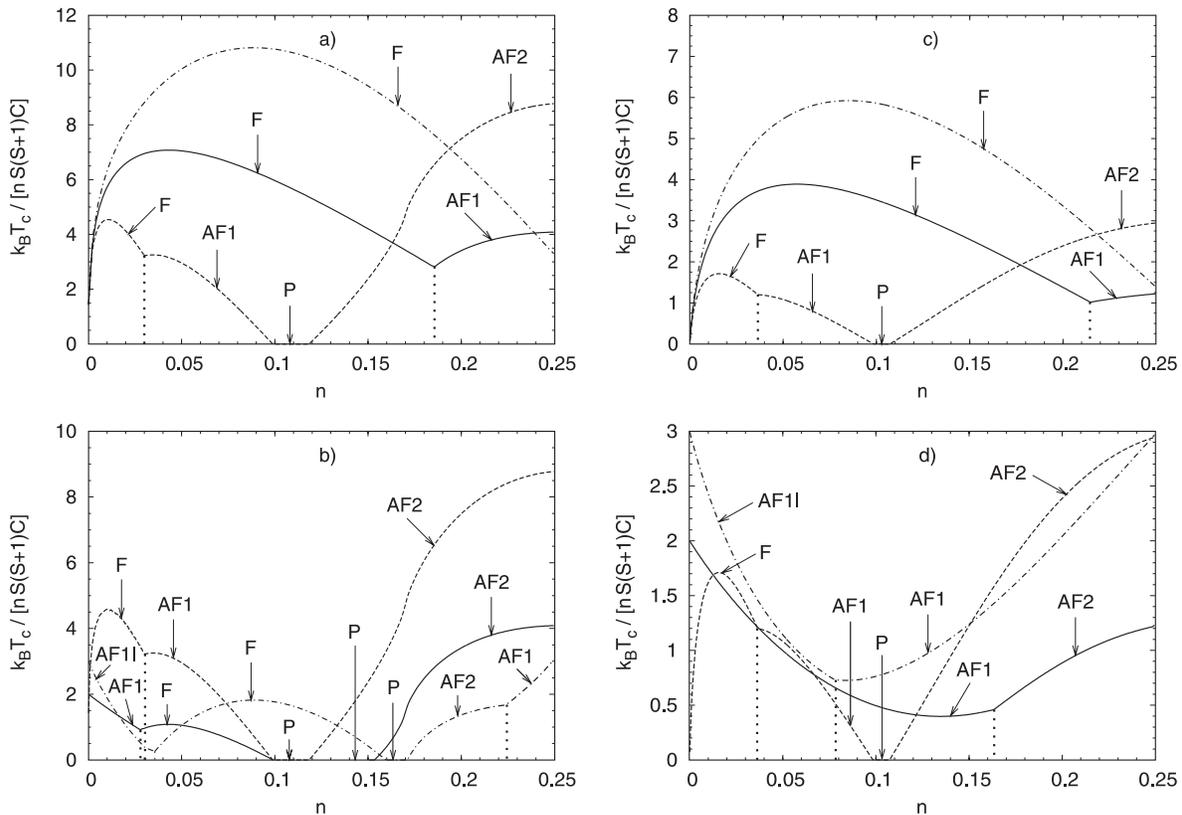}
\caption{\label{fig:fig2}MFA reduced critical temperature vs. charge carriers concentration, for different values of parameters modifying the RKKY interaction: $\lambda\to\infty$, $J^{\mathrm{AF}}=0$ (a), $\lambda\to\infty$, $J^{\mathrm{AF}}/C=-0.5$ (b), $\lambda/a=1.0$, $J^{\mathrm{AF}}=0$ (c), $\lambda/a=1.0$, $J^{\mathrm{AF}}/C=-0.5$ (d) and for different structural correlations: $\alpha_1=0$ (solid line), $\alpha_1=-1.0$ (dashed line) and $\alpha_1=0.5$ (dashed-dotted line).}
\end{figure*}
Fig.~1(a) is prepared for $\lambda \to \infty$ and  $J^{\mathrm{AF}}=0$, i.e., for the "pure" RKKY interaction. The ferromagnetic (F) phase is present for the low concentration $n$ only, but for all values of  $\alpha_1$. The positive $\alpha_1$  values (with structural clustering) enlarges the stability of the F phase. On the other hand, for $\alpha_1<0$ the antiferromagnetic phases become more favourable, and even the paramagnetic (P) phase can occur in some small restricted area. There are three stability regions for AF1-phase in Fig.~1(a). The rest of phases occur in the single regions only. In Fig.~1(b) the antiferromagnetic NN interaction, with the value $J^{\mathrm{AF}}/C=-0.5$, is included. As a result the F-phase area is strongly reduced, whereas the antiferromagnetic phases are extended. There are four stability regions for AF1-phase, and two for each AF1I and AF2. The new areas (AF1, AF1I and AF2) appear for the lowest values of $n$. Moreover, the P-phase becomes stable over all values of $\alpha_1$ in some narrow range of concentrations $n$.  In Fig.1(c) we assume $J^{\mathrm{AF}}=0$ and $\lambda/a =1$, which represents a relatively strong charge carriers localization. The topology of this diagram is somewhat similar to Fig.~1(a), with some differences in the shapes of the particular lines. Two areas of antiferromagnetic AF1-phases (one for strong negative and other for strong positive $\alpha_1$-parameters) are smaller than the corresponding areas in Fig.~1(a), i.e., without the RKKY damping. In the last part (Fig.~1(d)) both non-zero values of $J^{\mathrm{AF}}/C=-0.5$ and  $\lambda/a =1$ were assumed. As a result, the F and P phases have been reduced to the very small areas. It is seen that for the positive structural clustering, with $\alpha_1 > 0$, only the antiferromagnetic phases are in favour in Fig.~1(d). There are two AF1 and also two AF1I areas which are stable for these phases in that figure. The AF1 region has remarkably increased, whereas the AF2 phase has been limited mainly to the negative SRO parameters. \\

From the analysis of Figs.~1(a)-(d) some general conclusions can be drawn. The SRO parameter exerts a strong influence on the phase diagram, but not for all values of $n$ on the equal footing. The greatest changes are observed for small $n$ values, i.e., for strong dilution. Both $J^{\mathrm{AF}}$ and  $\lambda$ parameters, which modify the RKKY interaction, have also important meaning, which is manifested in Fig.~1 by remarkable divergences between various parts ((a)-(d)) of the figure. All the phase boundaries presented in Fig.1 constitute the discontinuous (1st order) phase transitions.\\

\subsection{Critical temperature}
The transition temperature from spontaneously ordered to disordered state is calculated on the basis of Eq.~(\ref{eq27}). In connection with the ground-state phase diagrams, the most important changes are predicted for small $n$-values; therefore we will present the results of critical temperature calculations in the range of $n\leq 0.25$. In Fig.~2, in the parts (a)-(d), the reduced critical temperature $T_{\mathrm{c}}/\left[nS(S+1)C\right]$ vs. $n$ is presented, for the same parameters of modified RKKY interaction as in Fig.~1 in the parts (a)-(d), respectively. The reduced temperature provides the results independent of the spin magnitude $S$ and shows a non-linear dependency of  $T_{\mathrm{c}}$ upon $n$.  On each phase diagram (2(a)-2(d)) the three curves are presented: solid, dashed and dashed-dotted, for $\alpha_1=0$, $-1 $ and $0.5$, respectively. Thus, the temperature phase diagrams in Fig.~2 are prepared as the horizontal cross-sections for the corresponding ground-state phase diagrams presented in  Fig.~1. The values of $\alpha_1$ have been chosen on the basis that enables to compare the calculations for the extremal SRO-parameters ($\alpha_1=-1$ and $\alpha_1=0.5$) with that for the absence of SRO ($\alpha_1=0$). The critical temperatures for each specific phase in Fig.~2 are pointed by the arrows with the phase symbols. The vertical (dotted) lines indicate the discontinuous phase boundaries, from P-phase down to the ground state, and their positions upon $n$ are in agreement with Fig.~1.\\

Some interesting features observable in Fig.~2 should be emphasized. In particular, in Fig.~2(a), for the unmodified RKKY interaction, it is seen that the positive clustering increases the Curie temperature and makes the region of F-phase wider. On the other hand, the negative structural correlations reduce both the Curie temperatures and the width of the ferromagnetic region. For the negative
SRO parameters, with increase of $n$ a relatively high N\'{e}el temperatures for AF2 phase can be achieved. 

For the antiferromagnetic NN interaction with the value $J^{\mathrm{AF}}/C=-0.5$ (Fig.~2(b)) the situation seems somewhat more complicated. The ferromagnetic phase for $\alpha_1=0$ and $\alpha_1=0.5$ is weaker than for the corresponding cases in Fig.~2(a); however, for $\alpha_1=-1$ it has a very similar phase boundary. The characteristic features are three paramagnetic gaps (one for each curve) occurring in accordance with Fig.~1(b). Also, for the small values of $n$ the new phases AF1 and AF1I are perceptible, with their N\'{e}el temperatures decreasing with increase of magnetic impurity (and charge) concentration $n$. In Fig.2(c) the influence of the charge carriers localization (with  $\lambda/a=1$) on the Curie temperature is demonstrated. When compared with Fig.~2(a), 
a remarkable reduction of all critical temperatures is apparent although the influence of SRO on the phase boundary shapes is comparable for both figures (1(a) and 2(c)) over the presented range of $n$. In Fig.~2(d), when both non-zero parameters ($J^{\mathrm{AF}}/C=-0.5$ and $\lambda/a=1$) are taken into account, the ferromagnetic phase is present only for $\alpha_1=-1$. For all curves in Fig.~2(d) a minimum of $T_{\mathrm{c}}$ is present in the region of concentrations for $n\approx 0.07-0.15$.\\

All the phase transitions from ordered to paramagnetic state, shown in Fig.~2, are continuous (2nd order) ones. However, for the vertical (dotted) lines, the perpendicular phase transitions (for constant temperature) between different ordered states are of discontinuous character, starting from P-phase down to the ground state. The points of junction, where two different 2nd order phase transition lines merge with another 1st order phase transition line are usually called the bicritical points. It can be noted that with the change of concentration $n$ the phase transition temperatures can either increase or decrease, and the changes are strongly non-linear. This effect depends not only on the range of concentration $n$ but on the SRO parameter as well.\\

\subsection{Magnetic frustration}
\begin{figure}
\includegraphics[scale=0.67]{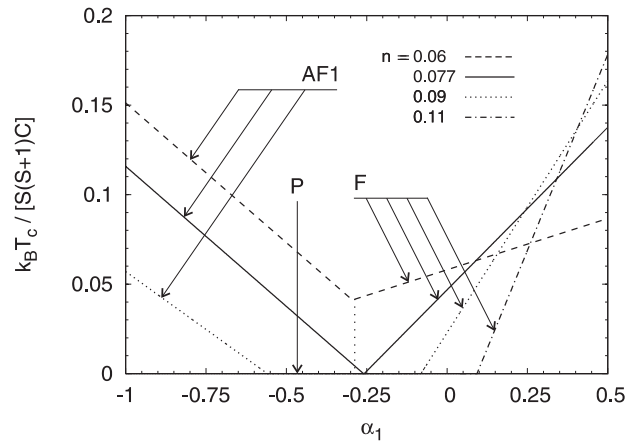}
\caption{\label{fig:fig3}MFA critical temperature in the vicinity of a selected triple point, $\left(n,\alpha_1\right)=\left(0.077,-0.25\right)$ from Fig.~1(b), vs. Warren-Cowley parameter $\alpha_1$, for various charge carriers concentrations $n$.}
\end{figure}
\begin{figure}
\includegraphics[scale=0.67]{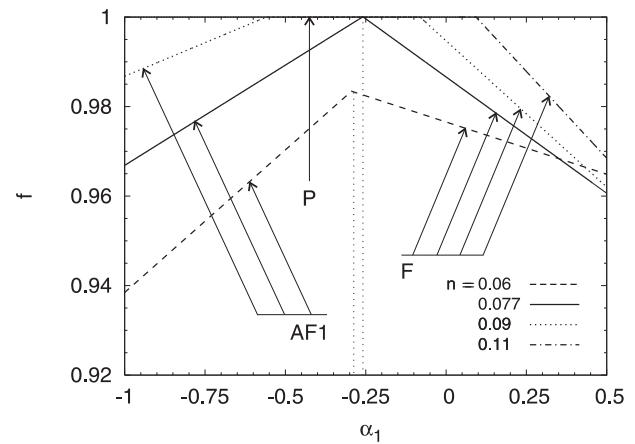}
\caption{\label{fig:fig4}Degree of frustration $f$ in the vicinity of a selected triple point with the same parameters as in Fig.~3, vs. Warren-Cowley parameter $\alpha_1$, for various charge carriers concentrations $n$.}
\end{figure}
\begin{figure*}
\includegraphics[scale=0.67]{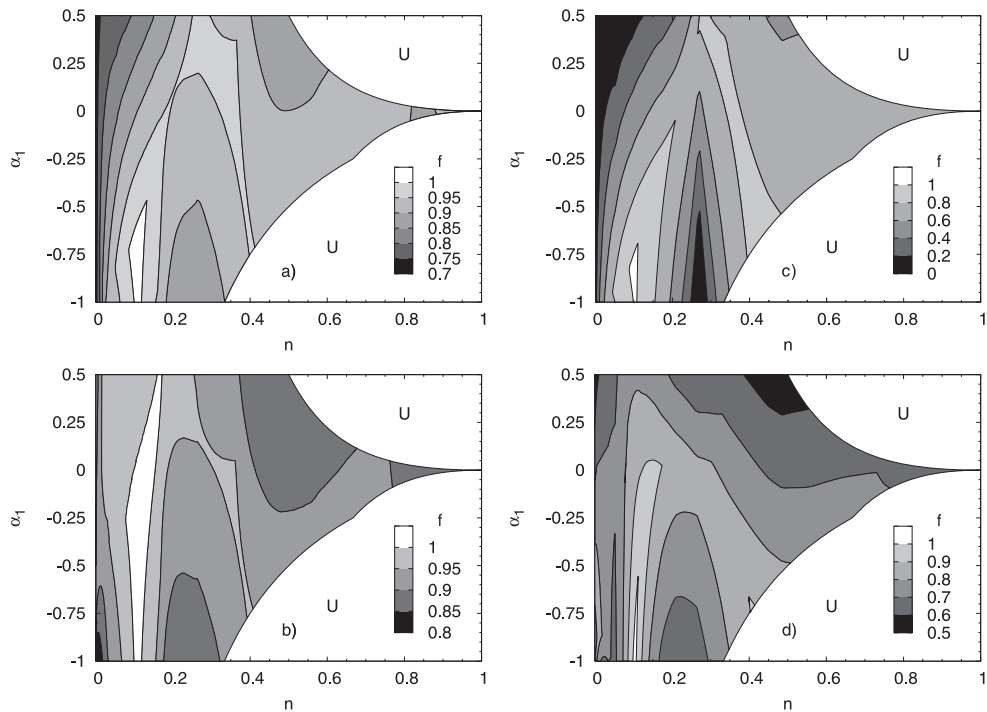}
\caption{\label{fig:fig5}Distribution of $f$-parameter over $\left(n,\alpha_1\right)$-plane for $J^{\mathrm{AF}}/C=0$ and $\lambda \to \infty$. The contour lines connect points with the same value $f$ of magnetic frustration. Note the differences in $f$ scale in the diagrams.}
\end{figure*}
Magnetic frustration is unavoidable for systems with the long-range oscillatory interaction. The presence of frustration leads to the increase of magnetic energy and decrease of the critical temperature. In order to define quantitatively the degree of frustration, we introduce an "ideal" ground-state energy $E_0$ in the form:
\begin{equation}
E_0=-\sum_{\left\langle i,j \right\rangle}^{}{\left|J\left(r_{ij}\right)\right|}=-\frac{N}{2}\sum_{k}^{}{z_k\left|J\left(r_k\right)\right|}.
\end{equation}
$E_0$ is the lowest magnetic energy which would be achieved in the hypothetical case, when no frustrations were present in the system. However, the real internal energy in the frustrated ground state, $E_g=\left(H\right)_{h=0}$, is given by the formulas (\ref{eq7}) and (\ref{eq8}) for the external field $h=0$. Thus, we can define the degree of frustration $f$ in the ground state as follows:
\begin{equation}
f=\left|\frac{E_g-E_0}{E_0}\right|.
\end{equation}

The value of $f$ comes from the range $\left<0; 1\right>$, where $f=0$ is for the "ideal" unfrustrated system, and $f=1$ stands for the paramagnetic phase with $E_g=0$.\\

In order to demonstrate a correlation of the degree of frustration with the phase transition temperature, first in Fig.~3 we present the four reduced $T_{\mathrm{c}}$-curves for different $n$ parameters, when $J^{\mathrm{AF}}/C=-0.5$ and $\lambda \to \infty$. The choice of $n$-parameters ($n=0.06,\, 0.077,\, 0.09$ and $0.11$) enables the vertical scan of Fig.1(b) upon $\alpha_1$, in the vicinity of the triple point which has the coordinates $\left(n,\alpha_1\right)=\left(0.077,-0.25\right)$. In this triple point in Fig.~1(b) the three phases: F, AF1 and P do coexist. The temperature studies in Fig.~3 show both N\'{e}el and Curie temperatures, as well as the existence of paramagnetic (P) phase, where $T_{\mathrm{c}}=0$.  On the other hand, in Fig.~4 the degree of magnetic frustration $f$ in the ground state is presented, for the same set of parameters ($J^{\mathrm{AF}}, \lambda, n$), and abscissa axis $\alpha_1$, as in Fig.~3. By comparison of Figs.~3 and 4 it is seen that the increase (decrease) of the critical temperature is accompanied by the decrease (increase) of the degree of frustration in the ground state, respectively. Both $T_{\mathrm{c}}$ and $f$ are the linear functions of $\alpha_1$, showing a non-smooth behaviour upon $\alpha_1$ at the 1st order phase transitions. A relatively high degree of frustration in Fig.~4 is remarkable, with its maximum in the vicinity of the triple point. It is worth noticing that at the triple point the chemical potentials of all coexisting phases are equal, and, because of the presence of P-phase, $f=1$.\\

In order to see how magnetic frustrations are distributed over $\left(n,\alpha_1\right)$-plane, in Fig.~5 the $f$-parameter is presented in the form of iso-$f$ contour lines. The interaction parameters in Fig.~5 amount to ($J^{\mathrm{AF}}=0$, $\lambda \to \infty$) (a), ($J^{\mathrm{AF}}/C=-0.5$, $\lambda \to \infty$) (b), ($J^{\mathrm{AF}}=0$, $\lambda/a=1$) (c) and ($J^{\mathrm{AF}}/C=-0.5$, $\lambda/a=1$) (d), i.e., are the same as in Figs.~1(a)-(d), respectively. The unphysical range of SRO-parameter is again denoted by U. The iso-$f$ contour lines connect the points with the same value of the frustration parameter $f$, according to the figures legend. For two figures ((a) and (b)) a relatively high values of $f$ can be noticed. 
On the other hand, in Fig.~5(c) a remarkable reduction of the $f$-parameter can be observed, when compared with Fig.~5(a). It is worthy to remind that respective phase diagrams (Fig.~1(a) and Fig.~1(c)) have a similar topology and have been prepared with the difference in $\lambda$-parameter only. Thus, it is hard to escape the obvious conclusion that the damping of RKKY interaction reduces the degree of frustration. In contrast, a presence of NN antiferromagnetic interaction on the FCC lattice increases the frustrations, which are then unavoidable even for the neighbouring spins.
A small white areas in Figs.~5(a)-(d) with $f=1$ correspond to the paramagnetic phases as seen in Figs.~1(a)-(d), respectively. However, the remaining contour lines of $f$ do not reflect the shapes of the ground-state phase boundaries. The distribution of $f$-parameter vs. $\alpha_1$ shows that the SRO has a strong influence on the frustration. However, this effect depends simultaneously on the magnetic dilution $n$.\\

\section{Conclusion}

The magnetic phase diagrams for the diluted FCC lattice with modified RKKY interaction have been studied, with SRO parameter taken into account. The formalism necessary for structural averaging in the pair approximation has been presented in the Appendix. As a result, the physical range of Warren-Cowley parameter has been established. In the theoretical part, the general statistical-thermodynamical method has been presented. On the basis of the above formalism, the phase diagrams and other magnetic properties can be studied in the external field $h$. The method can also be adopted for other kinds of localized spins Hamiltonians. The numerical calculations show that the SRO parameter $\alpha_1$ has a remarkable influence on the phase diagrams, both in the ground state and for the temperature dependency, especially for low $n$. Both positive  and negative structural ordering parameters have been considered. 
A comparison of the results with those for the virtual crystal approximation can be easily made by assuming $\alpha_1=0$.
The temperature phase diagrams contain the vertical phase boundaries, which represent 1st order phase transitions for $T\geq 0$. The diluted system with RKKY interaction is mostly a very frustrated one, and the frustration parameter $f$ depends both on the SRO and the parameters of modified interaction. The inverse correlation between the degree of frustration and the critical temperature has been established. One of the novelty of the presented method lies in incorporating the Warren-Cowley parameter into the magnetic model, which can be easily extended to other systems, where the structural clustering can play a role in magnetic phenomena.\\

\begin{acknowledgments}
This work was supported by the European Social Fund and Budget of State implemented under the Integrated Regional 
Operational Program, action 2.6, project GRRI-D.
\end{acknowledgments}

\appendix*

\section{Atomic short range order in diluted systems}

We consider a system of localized spins on the crystalline lattice. In such a system, the spin dilution can be conveniently described by means of site occupation operators $\xi_i$~\cite{edwards} for each site, possessing the eigenvalues 0 (the corresponding $i$-th site is empty) and 1 (the $i$-th site is occupied by the magnetic impurity ion).  The probability  distribution of the eigenvalues of the single-site occupation operator is as follows (see ex. Ref.~\onlinecite{bonfim}):
\begin{equation}
\label{eq:onesite}
p\left(\xi_i\right)=n\delta\left(\xi_i-1\right)+\left(1-n\right)\delta\left(\xi_i\right).
\end{equation}
where $\delta\left(\xi_i-1\right)$ and  $\delta\left(\xi_i\right)$  are the Kronecker's deltas.
It can be verified that $\sum_{\xi_i=0,1}^{}{p\left(\xi_i\right)}=1$, thus the distribution is nomalized.   

The configurational average of the operator $\xi_i$ equals
\begin{equation}
\left\langle\xi_i\right\rangle_r=\sum_{\xi_i=0,1}^{}{\xi_i p\left(\xi_i\right)},
\label{eq:average_xi}
\end{equation}
so that it is the occupation probability for a single site (i.e. average number of magnetic impurities per site). The average $\left\langle\xi_i\right\rangle_r$ is independent on the $i$-th site location, which comes from the fact that the distribution (\ref{eq:onesite}) is valid for the whole sample. \\

Let us consider a general probability distribution of the occupation of a pair of sites $i$ and $j$:
\begin{eqnarray}
p\left(\xi_i,\xi_j\right)&=&p^{00}_{ij}\delta\left(\xi_i\right)\delta\left(\xi_j\right)+p^{01}_{ij}\delta\left(\xi_i\right)\delta\left(\xi_j-1\right)\nonumber\\
&&+p^{10}_{ij}\delta\left(\xi_i-1\right)\delta\left(\xi_j\right)+p^{11}_{ij}\delta\left(\xi_i-1\right)\delta\left(\xi_j-1\right).\nonumber\\
\label{eq:twosite}
\end{eqnarray}
The number $p^{AB}_{ij}$ is the probability of the event $AB$ ($A,B=0,1$) for the pair of lattice sites $i$ and $j$.

The normalization requires:
\begin{equation}
p^{00}_{ij}+p^{01}_{ij}+p^{10}_{ij}+p^{11}_{ij}=1.
\label{eq:norm}
\end{equation}
The simplest approach to the description of the diluted system, so called Virtual Crystal Approximation (VCA), relies on the assumption that the impurity ions are distributed randomly in the lattice sites, thus they are uncorrelated. Then the occupations of the specified sites are statistically independent events and the probability distribution (\ref{eq:twosite}) takes a product form $p\left(\xi_i,\xi_j\right)=p\left(\xi_i\right)p\left(\xi_j\right)$, so $p^{00}_{ij}=\left(1-n\right)^2$, $p^{01}_{ij}=p^{10}_{ij}=n\left(1-n\right)$ and  $p^{11}_{ij}=n^2$.

In such a situation, the only parameter describing the distribution of magnetic impurities is $n$.

However, VCA does not include the possible Short Range Ordering (SRO) in the diluted system, which takes place when the occupations of specific sites are not independent events, implying then $p\left(\xi_i,\xi_j\right)\neq p\left(\xi_i\right)p\left(\xi_j\right)$. In particular when $p\left(\xi_i,\xi_j\right)>p\left(\xi_i\right)p\left(\xi_j\right)$ we deal with clustering. The existence of SRO stems from the interactions between the impurity ions (especially the coulombic ones). 

The general form of the two-site probability distribution (\ref{eq:twosite}) must be reducible to one-site distributions:
\begin{subequations}
\begin{equation}
\sum_{\xi_i=0,1}^{}{p\left(\xi_i,\xi_j\right)}=p\left(\xi_j\right)
\label{eq:reduction_i}
\end{equation}
\begin{equation}
\sum_{\xi_j=0,1}^{}{p\left(\xi_i,\xi_j\right)}=p\left(\xi_i\right).
\label{eq:reduction_j}
\end{equation}
\end{subequations}
The above conditions, together with Eq.~\ref{eq:onesite}, yield:
\begin{subequations}
\begin{eqnarray}
p^{00}_{ij}+p^{10}_{ij}&=&n\nonumber\\
p^{01}_{ij}+p^{11}_{ij}&=&1-n
\label{eq:reduction_i2}
\end{eqnarray}
\begin{eqnarray}
p^{00}_{ij}+p^{01}_{ij}&=&n\nonumber\\
p^{10}_{ij}+p^{11}_{ij}&=&1-n.
\label{eq:reduction_j2}
\end{eqnarray}
\end{subequations}
Using Eq.~\ref{eq:norm} we obtain:
\begin{equation}
\label{eq:prob}
p^{00}_{ij}=1-2n+p^{11}_{ij}\qquad p^{01}_{ij}=p^{10}_{ij}=n-p^{11}_{ij}.
\end{equation}
Thus, the general two-site probability distribution can be parametrized independently by $n$ and $p^{11}_{ij}$.

The average of the pair occupation operator reads:
\begin{equation}
\left\langle\xi_i\xi_j\right\rangle_r=\sum_{\xi_i,\xi_j=0,1}^{}{\xi_i\xi_j p\left(\xi_i,\xi_j\right)}=p^{11}_{ij},
\end{equation}
thus it equals the probability of occupation of both sites $i$ and $j$ simultaneously. For VCA, the average factorizes: $\left\langle\xi_i\xi_j\right\rangle_r=\left\langle\xi_i\right\rangle_r\left\langle\xi_j\right\rangle_r^2$ due to the product form of $p\left(\xi_i,\xi_j\right)$. 

Let us write the site occupation operator as:
\begin{equation}
\xi_i=\left\langle\xi_i\right\rangle_r+\delta\xi_i,
\end{equation}
where $\left\langle\delta \xi_i\right\rangle_r=0$ to fulfill Eq~\ref{eq:average_xi}.
Thus, in a general case we obtain:
\begin{equation}
\left\langle\xi_i\xi_j\right\rangle_r^2=n^2+\left\langle\delta\xi_i\delta\xi_j\right\rangle_r.
\end{equation}
Let us denote $\left\langle\delta\xi_i\delta\xi_j\right\rangle_r=\Delta_k$. The above expression describes isotropic correlations, i.e. the value $\left\langle\xi_i\xi_j\right\rangle_r$ depends solely on the distance $r_k$ between sites $i$ and $j$ (the $j$-th site lies on the $k$-th coordination zone of the $i$-th site). In particular, we have $\Delta_k=0$ in absence of SRO, for VCA. In the above notation the probabilities are:
\begin{eqnarray}
\label{eq:prob2}
p_{ij}^{00}&=&1-2n+n^2+\Delta_k \qquad p^{01}_{ij}=p^{10}_{ij}=n-n^2-\Delta_{k}\nonumber\\
p_{ij}^{11}&=&n^2+\Delta_k. 
\end{eqnarray}
One can also write the conditional probabilities:
\begin{eqnarray}
\label{eq:cond}
 p_{ij}^{0|1}&=&\frac{n-n^2-\Delta_k}{1-n}\qquad p_{ij}^{0|0}=1-\frac{n-n^2-\Delta_k}{1-n} \nonumber\\
p_{ij}^{1|1}&=&\frac{n^2+\Delta_k}{n} \qquad p_{ij}^{1|0}=1-\frac{n^2+\Delta_k}{n},
\end{eqnarray}
where $p_{ij}^{A|B}$ is the conditional probability of an event $B=0,1$ for the site $j$ under condition that an event $A=0,1$ occurred for the site $i$. It can be verified that the above probabilities obey Bayes theorem as well as complete probability theorem. 

The possible physical range of $\Delta_k$ follows from the interpretation of $p^{00}_{ij}$, $p^{01}_{ij}$, $p^{10}_{ij}$ and $p^{11}_{ij}$ as probabilities, obeying $0\leq p_{ij}^{AB}\leq 1$. Using Eq.~\ref{eq:prob2} we arrive at the set of inequalities:
\begin{eqnarray}
\label{eq:ineq1}
-n^2+2n-1&\leq&\Delta_k\leq -n^2+2n\nonumber\\
-n^2+n-1&\leq&\Delta_k\leq -n^2+n\nonumber\\
-n^2&\leq&\Delta_k\leq -n^2+1,
\end{eqnarray}
which must hold for every $k$.

There exists another condition to impose on the parameters $\Delta_k$. Let us consider the operator of the total number of occupied pairs, $\frac{1}{2}\sum_{i,j\neq i}^{}{\xi_i\xi_j}$, 
with the average:
\begin{equation}
\frac{1}{2}\sum_{i,j\neq i}^{}{\left\langle\xi_i\xi_j\right\rangle_r}=\frac{1}{2}N\sum_{k}^{}{z_k\left(n^2+\Delta_k\right)}.
\label{eq:pairs}
\end{equation}
In the equation above, $z_k$ denotes the number of lattice sites on the $k$-th co-ordination zone. 
If we deal with a constant number of impurity ions in the lattice, then the total number of occupied pairs is also constant and does not depend on the SRO existence. As mentioned above,  $\Delta_k\equiv 0$ in absence of SRO. Therefore, by comparison of Eq.~\ref{eq:pairs} taken for the absence of SRO and for arbitrary choice of impurity correlations we obtain a constraint:
\begin{equation}
\label{eq:constraint}
\sum_{k}^{}{z_k\Delta_k}=0.
\end{equation}
Let us restrict to the situation, when SRO is limited to nearest- and next-nearest neighbours of each site, i.e. when $\Delta_1\neq 0 $ and $\Delta_2\neq 0$, while $\Delta_k=0$ for $k>2$. Notice that the constraint (\ref{eq:constraint}) excludes the possibility that only a single value of $\Delta_k$, for instance for n.n, is non-zero. Therefore, the situation we selected involves minimal number of $\Delta_k$ parameters. Since the constraint gives $ \Delta_2=-\left(z_1/z_2\right)\Delta_1$ the distribution of impurities is then described by two independent numbers $n$ and $\Delta_1$. 

Writing the inequalities (\ref{eq:ineq1}) for $\Delta_1$ and for $\Delta_2$ expressed by $\Delta_1$, we arrive at the allowed range of the parameter $\Delta_1$, which has to fulfill the following 12 inequalities:
\begin{eqnarray}
\label{eq:ineq2}
-n^2+2n-1&\leq&\Delta_1\leq -n^2+2n\nonumber\\
-n^2+n-1&\leq&\Delta_1\leq -n^2+n\nonumber\\
-n^2&\leq&\Delta_1\leq -n^2+1\nonumber\\
\frac{z_2}{z_1}\left(n^2-2n\right)&\leq&\Delta_1\leq \frac{z_2}{z_1}\left(n^2-2n+1\right)\nonumber\\
\frac{z_2}{z_1}\left(n^2-n\right)&\leq&\Delta_1\leq \frac{z_2}{z_1}\left(n^2-n+1\right)\nonumber\\
\frac{z_2}{z_1}\left(n^2-1\right)&\leq&\Delta_1\leq \frac{z_2}{z_1}n^2.
\end{eqnarray}
\begin{figure}
\includegraphics[scale=0.71]{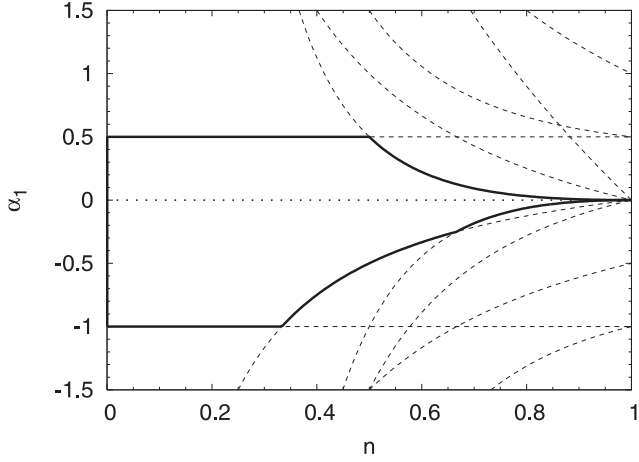}
\caption{\label{fig:fig6}The allowed range of Warren-Cowley parameter for the first co-ordination zone of the fcc lattice. The dashed lines (12 in total) are the limits obtained from the inequalities (\ref{eq:ineq2}). The thick solid line encloses the allowed range of $\alpha_1$ for various $n$. The dotted line for $\alpha_1=0$ corresponds to vanishing SRO.}
\end{figure}

The existence of SRO in diluted systems can also be described conveniently by means of the Warren-Cowley (W-C) parameters $\alpha_{ij}$,~\cite{cowley1} defined as:
\begin{equation}
\alpha_{ij}=\frac{\left\langle\xi_i\xi_j\right\rangle_r-\left\langle\xi_i\right\rangle_r\left\langle\xi_j\right\rangle_r}{\left\langle\xi_i\right\rangle_r\left\langle\xi_j\right\rangle_r}.
\end{equation}
These parameters can be equivalently given in the form $\alpha_{ij}=-\left(1-p^{1|1}_{ij}/n\right)$, where $p^{1|1}_{ij}=p^{11}_{ij}/n$ is a conditional probability of occupying the $j$-th site if the $i$-th site is occupied. 

In our notation the W-C parameter for $k$-th coordination zone is given by:
\begin{equation}
\alpha_k=\Delta_k/n^2
\end{equation}
and the configurational average of pair occupation operator reads:
\begin{equation}
\label{eq:correlations}
\left\langle\xi_i\xi_j\right\rangle_r^2=n^2\left(1+\alpha_k\right).
\end{equation}
The allowed range of W-C parameter $\alpha_1$ values for $\Delta_1$ fulfilling the inequalities~(\ref{eq:ineq2}) for fcc lattice is presented in the Fig.~\ref{fig:fig6}, where it is bounded by thick solid lines. In this range the physically possible correlations are given by Eq.~(\ref{eq:correlations}). For other lattices characterized by specific sets of $z_k$ numbers the range of W-C parameter requires separate calculations based on the inequalities (\ref{eq:ineq2}).


\end{document}